\documentclass[usenatbib]{mnras}
\usepackage{academicons}
\usepackage{mathptmx}
\usepackage{color}
\usepackage{xcolor}
\usepackage{graphics,graphicx,amssymb,amsmath}
\usepackage{hyperref}
\usepackage{academicons}
\usepackage{multirow}
\usepackage{soul}
\soulregister\citepp7
\soulregister\citept7
\soulregister\ref7

\newcommand{\orcid}[1]{\href{https://orcid.org/#1}{\textcolor[HTML]{A6CE39}{\aiOrcid}}}

\def\gtrsim{\lower 2pt \hbox{$\, \buildrel {\scriptstyle >}\over
{\scriptstyle \sim}\,$}}
\def\lesssim{\lower 2pt \hbox{$\, \buildrel {\scriptstyle <}\over
{\scriptstyle \sim}\,$}}

\def\hst{{\sl HST}}

\def\xmm{{\sl XMM-Newton}}

\def\chandra{{\sl Chandra}}

\def\HST{{\sl HST}}
\def\hst{{\sl HST}}

\def\xsb{PJ0116-24}
\def\xsa{PJ1336+49}
\def\xsc{PJ1053+60}
\def\sw{$\rm AGN_{\rm SW}$}
\def\fg{$\rm AGN_{\rm FG}$}

\def\approxlt{\lower.2em\hbox{$\buildrel < \over \sim$}}
\def\approxgt{\lower.2em\hbox{$\buildrel > \over \sim$}}
\def \ls   {\hbox{L$_{\odot}$}}

\def\sous{HyLIRGs}

\def\xp{$X_{\sc HMXB}$}
\def\ins{{\sl XMM-Newton}}

\date{Accepted XXX. Received YYY; in original form ZZZ}

\pubyear{2025}

\begin{document}
\title[]{The complicated nature of the X-ray emission from the field of  the strongly lensed hyperluminous infrared galaxy \xsc\ at $z$=3.549}
\author[Carlos Garcia Diaz$^{1}$]{Carlos Garcia Diaz$^{1}$\thanks{Contact e-mail:cgarciadiaz@umass.edu}\href{https://orcid.org/0009-0004-1123-001X}{\includegraphics[scale=0.7]{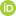}}, Q. Daniel Wang$^{1}$, Kevin C. Harrington$^{2,3,4,5}$ \href{https://orcid.org/0000-0001-5429-5762}{\includegraphics[scale=0.7]{f/ORCID-iD_icon_16x16.png}}, James D. Lowenthal$^{6}$,  \newauthor{Patrick S. Kamieneski$^{7}$ \href{https://orcid.org/0000-0001-9394-6732}{\includegraphics[scale=0.7]{f/ORCID-iD_icon_16x16.png}}}, Eric F. Jimenez-Andrade$^{8}$\href{https://orcid.org/0000-0002-2640-5917}{\includegraphics[scale=0.7]{f/ORCID-iD_icon_16x16.png}}, Nicholas Foo$^{7}$, Min S. Yun$^{1}$, Brenda L. Frye$^{9}$, \newauthor{Dazhi Zhou$^{10}$}\href{https://orcid.org/0000-0002-6922-469X}{\includegraphics[scale=0.7]{f/ORCID-iD_icon_16x16.png}}, Amit Vishwas$^{11}$\href{https://orcid.org/0000-0002-4444-8929}{\includegraphics[scale=0.7]{f/ORCID-iD_icon_16x16.png}}, Ilsang, Yoon$^{12}$,
Bel\'{e}n Alcalde Pampliega$^{2}$, Daizhong Liu$^{13}$, \& \newauthor{Massimo Pascale$^{14}$}\\
$^{1}$ Department of Astronomy, University of Massachusetts, Amherst, MA 01003, USA\\
$^{2}$ Joint ALMA Observatory, Alonso de C´ordova 3107, Vitacura, Casilla 19001, Santiago de Chile, Chile \\
$^{3}$ National Astronomical Observatory of Japan, Los Abedules 3085 Oficina 701, Vitacura 763 0414, Santiago, Chile \\
$^{4}$ European Southern Observatory, Alonso de C´ordova 3107, Vitacura, Casilla 19001, Santiago de Chile, Chile \\
$^{5}$ Instituto de Estudios Astrof´ısicos, Facultad de Ingenier´ıa y 455 Ciencias, Universidad Diego Portales, Av. Ej´ercito Libertador 441,
Santiago, Chile \\
$^{6}$ Department of Astronomy, Smith College, Northampton, MA 01063, USA\\
$^{7}$ School of Earth and Space Exploration, Arizona State University, Tempe, AZ 85287-6004, USA\\ 
$^{8}$ Instituto de Radioastronomía y Astrofísica, Universidad Nacional Aut\'{o}noma de M\'{e}xico, Antigua Carretera a P\'{a}tzcuaro \# 8701,\\ Ex-Hda. San Jos\'{e} de la Huerta, Morelia, Michoacán, M\'{e}xico C.P. 58089\\
$^{9}$ Department of Astronomy/Steward Observatory, 933 North Cherry Avenue, University of Arizona, Tucson, AZ 85721, USA\\
$^{10}$ Department of Physics and Astronomy, University of British Columbia, 6225 Agricultural Rd., Vancouver, V6T 1Z1, Canada \\
$^{11}$ Department of Astronomy, Cornell University, Space Sciences Building, Ithaca, NY 14853, USA\\
$^{12}$ National Radio Astronomy Observatory, 520 Edgemont Road, Charlottesville, VA 22903\\
$^{13}$ Purple Mountain Observatory, Chinese Academy of Sciences, 10 Yuanhua Road, Nanjing 210023, China\\
$^{14}$Department of Astronomy, University of California, Berkeley, CA 94720, USA\\
 }
\maketitle
\begin{abstract}
We present an analysis of \xmm\ X-ray observations of \xsc, a hyperluminous infrared galaxy (HyLIRG) at $z$=3.549 that is strongly lensed by a foreground group at $z$=0.837. We also present GNIRS spectroscopy confirming the presence of an active galactic nucleus (AGN) to the southwest of \xsc\ ($\rm AGN_{\rm SW}$) at $z_{SW}=1.373\pm0.006$. Using this redshift prior, we decompose the X-ray spectrum of \xsc\ into $\rm AGN_{\rm SW}$ and high-mass X-ray binary (HMXB) components from the HyLIRG. The HMXB component has an unusually high luminosity, $\sim 50$ times higher than calibration derived from local galaxies, and a characteristic photon index likely too flat to be caused by high-mass X-ray binaries at rest frame energies above a few keV. Our 2-D spatial decomposition also suggests a similarly high X-ray HMXB luminosity, although the limited spatial resolution prevents meaningful morphological constraints on the component. We conclude that the enhanced X-ray emission may only be explained by the presence of another AGN ($\rm AGN_{\rm FG}$) embedded in the foreground group lensing the \xsc\ system. The presence of $\rm AGN_{\rm FG}$ is further supported by the detection of a point-like radio continuum source that coincides with the brightest group galaxy (BGG) of the foreground lens. 
Our study demonstrates the limited capability of current X-ray observatories while highlighting the need for higher angular resolution observations to definitively characterize the nature of X-ray emission in distant, strongly lensed HyLIRGs.

\end{abstract}
\begin{keywords}
galaxies: active, galaxies, galaxies: high-redshift, galaxies,
gravitational lensing: strong, physical Data and Processes, galaxies:
starburst, galaxies, X-rays: binaries, resolved and unresolved sources
as a function of wavelength, X-rays: galaxies 
\end{keywords}

\section{INTRODUCTION}
\label{s:intro}

Studying the most extreme galaxies is vital to better understand the physical processes that affect the evolution and growth of all galaxies. Hyperluminous infrared galaxies (HyLIRGs) are some of the most intrinsically luminous objects in the universe that exhibit high infrared (IR) luminosities $> 10^{13}$ L$_{\odot}$ and are typically found at $z>$2 \citep{Rowan-Robinson2000,Blain2002,Casey2014}. Examining these sources through X-ray observations is crucial to learn how high-energy phenomena like Active Galactic Nuclei \citep[AGN;][]{Farrah2002,Ruiz2007} and high-mass X-ray binaries \citep[HMXBs;][]{Fragos2013}
affect galaxy growth. These processes emit X-rays through accretion with AGN emission caused by one supermassive black hole (SMBH) \citep{Padovani2017} and HMXB emission caused by combined emission from accretion of an HXMB population \citep{Mineo2012,Fragos2013,Riccio2023}. Although high energy processes like the AGN fraction in less luminous LIRGs have been measured to be $\gtrsim 30\%$ in some cases \citep[e.g.,][]{Zhang2024,Torres-Alba2018,Iwasawa2011,Veilleux1999}, HyLIRGs' high redshifts make X-ray observations difficult, as X-ray emission is typically very faint at $z>1$. A hypothetical AGN at $z=1.0$ emitting a power-law spectrum with a luminosity of $10^{43}$ erg s$^{-1}$ and a photon index of 2, would have an expected X-ray count rate $\sim5$ times less than if the AGN were at $z=0.5$. Therefore, observations at $z>1$ require a monumental amount of effort and time with current X-ray observatories like \chandra\ and \xmm. Consequently, X-rays from HyLIRGs have historically been lightly explored, leaving many unanswered questions regarding the high-energy processes in these objects. 

\par Only the brightest objects, such as quasars \citep{Padovani2017} or HyLIRGs that are expected to harbor rich well distributed HMXB populations due to their extreme SFRs, are feasible to observe \citep{Wang2024}. Nevertheless, previous high redshift X-ray observations typically only obtain $\sim 10$ counts \citep[e.g.,][]{Shankar2020}, requiring such studies to rely on stacking different sources to produce usable spectra that may yield differing results on the same datasets \citep[e.g.,][]{Fornasini2018,lambrides2020A}. One may find help from gravitational lensing -- nature's telescope. Despite the effective usage of lensing, which has been well demonstrated in other wavelengths from radio to IR observations \citep[e.g.,][]{myers2003,Bolton2006,Harrington2018,Berman2022,Kamieneski2023,Guilietti2024}, lensed X-ray observations have been limited for high-$z$ galaxies because of their expected low X-ray fluxes, which still require large observing programs of $\gtrsim$ 100 ks. Only with the combination of gravitational lensing, extreme systems, and massive X-ray programs can detailed analysis be performed on the high-energy phenomena of high redshift sources.  

Observing HMXB populations at $z>2$ provides a unique perspective on the energetic activity of a galaxy. Because HMXBs are short-lived with lifespans ranging from $\sim10^5 - 10^7$ yrs \citep{Tan2021}, X-ray detections of HMXBs are typically found only in active star-forming regions. The X-ray luminosity ($L_X$) - Star Formation Rate ($SFR$) relation has previously only been observed in HMXB populations from local galaxies ($z \lesssim 1$); one notable study on this topic is presented in \cite{Mineo2012,Mineo2014}. The relation shown in Eq. \ref{e:xsfr}~(referred to as the Mineo relation hereafter) directly relates the $SFR$ and $L_X$ of a source and can be used to estimate either parameter. Although such studies have grown our understanding the $L_X - SFR$ relation, these observations are limited by low SFRs of sources at $z\lesssim 1$. Thus, higher redshift studies via gravitational lensing are necessary to measure the $L_X - SFR$ relation of high star-forming galaxies.  

Recent X-ray detections presented in \cite{Wang2024} of HyLIRGs, \xsa, \xsb, and \xsc~at redshifts ranging $z= 2.12 - 3.55$, have been shown to yield enough counts ($\gtrsim 50$ counts per source) to perform X-ray spectral analysis. Those sources are selected from the Planck All-Sky Survey to Analyze Gravitationally-lensed Extreme Starbursts (PASSAGES) sample \citep{Harrington2016,Berman2022}, which consists of 30 dusty star-forming galaxies (DSFGs). Most of the sources in the sample exhibit extremely high apparent IR luminosities ($\gtrsim 10^{14}$ \ls) and are gravitationally lensed. That sample has been extensively studied by using a large variety of observatories [e.g. Hubble Space Telescope (\hst), Large Millimeter Telescope (LMT), Atacama Large Millimeter Array (ALMA), etc.]. These multiwavelength observations in combination with lensing provide an excellent testbed to study various physical properties of the most extreme star-forming galaxies. Before the \xmm\ observations, the targets'  radio fluxes were mostly consistent with star formation \citep{Kamieneski2023} suggesting no clear evidence of AGN embedded in the lensed galaxies or foreground lensing systems. Therefore, the \xmm\ targets' foreground lens galaxies or groups were expected to emit low X-ray fluxes, allowing the study to focus on the emission from the lensed background sources. \cite{Wang2024} shows that the ratios of $L_x$ to $SFR$ in HyLIRGs appear to be systematically higher than those observed locally \citep{Mineo2012,Mineo2014}. In this paper and in \cite{Wang2024}, we use a simplified parameter, \xp, to reference the ratio of $L_X/SFR$ from the Mineo relation:
\begin{equation}
    X_{HMXB} \approx 3.9 \times 10^{39} ({\rm erg~s^{-1}}/{\rm~M_{\odot}~yr^{-1}}).
\label{e:xsfr} 
\end{equation}
Here, the \xp\ parameter is a multiplicative constant derived from the expected X-ray emission from the Mineo relation. An \xp\ $\sim 1$ would indicate we observe an $L_X$ that is consistent with the Mineo relation. We use this local calibration as a benchmark to compare our high-$z$ \xmm\ measurements of the $L_X - SFR$ relation.

\begin{figure*}
\centerline{
\includegraphics[width=1.0\linewidth,angle=0]{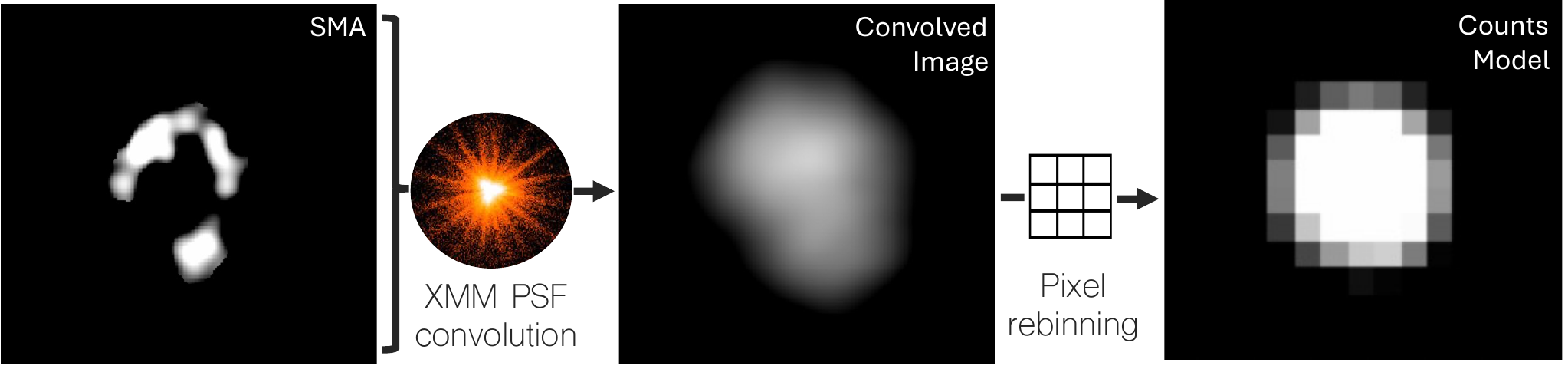}
}
\caption{Illustration showing the procedure to produce the \xp~ 2D image model. The original masked SMA image is convolved with the \xmm\ PSF. The convolved image is then rebinned to match the pixel coordinates of the \xmm\ image to produce the model, with each pixel value being the expected counts from the HMXB population. 
}
\label{f:2d-pro}
\end{figure*}

This paper describes a detailed follow-up study of  \xsc\ -- one of the three previously mentioned PASSAGES sources with \xmm\ observations. The background source, \xsc, is gravitationally lensed by a foreground galaxy group at $z$=0.837 that magnifies the DSFG by a factor of 24, as measured from the state-of-the-art Light Traces Mass (LTM) lens model \citep{Frye2019,Wang2024}. A 6 GHz source detection in VLA data corresponding with the position of the Brightest Group Galaxy (BGG) at (R.A./Dec. J2000)
10:53:22.707 / +60:51:46.157 in \hst\ images suggests the presence of an AGN (referred to as foreground AGN or $\rm AGN_{\rm FG}$ hereafter). Additionally, the \xmm\ study in \cite{Wang2024} shows that two of the three sources, including \xsc, exhibit an excess of hard X-ray emission ($>$ 2 keV), suggesting the existance of another AGN (referred to as southwest AGN or $\rm AGN_{\rm SW}$ hereafter). Given that X-ray observations are among the few robust tracers of AGN in HyLIRGs \citep[e.g.,][]{Ivison2019}, the rate of occurrence of AGN embedded in HyLIRGs has previously been difficult to assess because of the faintness of X-ray emission at $z\gtrsim $2. Mid-IR studies suggest that AGN could be present in $\sim 25 - 50 \%$ of galaxies with a high $SFR$ \citep{Kirkpatrick2017}. \cite{Wang2024} speculate that AGN$_{SW}$ could be within $\sim60$ kpc southwest of \xsc\ or at a physically unrelated redshift that may contaminate the integrated X-ray flux. Given the southwest AGN's close proximity, the AGN emission must be carefully handled to measure the $L_X-SFR$ relation accurately. 

To assess the contributions of $\rm AGN_{\rm SW}$ and confirm its spectroscopic redshift, we carried out deep K-band spectroscopic observations using GNIRS that are discussed in \S~\ref{s:gemini} and \S~\ref{s:gspec} Beyond the spectral analysis previously presented in \cite{Wang2024}, we conduct 2D spatial modeling to decompose the AGN and galaxy contributions to the observed X-ray emission of the system in \S~\ref{s:2dspat} and \S~\ref{s:2dres}. This provides a novel method of measuring the \xp\ value that is fully independent of the spectral modeling. Furthermore, the GNIRS observations permit a more detailed spectral analysis with a more robust AGN spectral model compared to the work in \cite{Wang2024}. Overall, our analysis presented here provides one of the most detailed modern X-ray studies of a galaxy at $z>3$.

\par The main goal of this paper is to disentangle the three components of this system: $\rm AGN_{\rm SW}$, $\rm AGN_{\rm FG}$, and \xsc. To analyze this complex scenario, we first measure the redshift of $\rm AGN_{\rm SW}$ to see if this source is a part of our HyLIRG system at $z=3.549$ or a contaminant interloper at a different redshift. This process is described in the Gemini observations and data reduction \S~\ref{s:gemini}. We then incorporate the redshift of \sw\ into the X-ray spectral model as described the spectral data \S~\ref{ss:d-spec} and modeling \S~\ref{ss:d-spec-modeling}. We then present our results and discuss their implications in \S~\ref{s:res}. We summarize our results and conclusions in \S~\ref{s:con}. 
Throughout this paper, all images are oriented north up and east to the left, and we assume a Kroupa Initial Mass Function and the $\Lambda$CDM cosmology\footnote{\url{https://astro.ucla.edu/~wright/CosmoCalc.html}} with {H\textsubscript{0}} = 69.6 ${\rm~km~s^{-1}~Mpc^{-1}}$, {$\Omega$\textsubscript{M}} = 0.286 and {$\Omega$\textsubscript{$\Lambda$}} = 0.714.

\section{Gemini Observations and Data Reduction}\label{s:gemini}

To accurately measure the X-ray emission of \xsc\, we must first determine the redshift of the southwest AGN. We determine the position of this AGN to be at (R.A./Dec. J2000) 10:53:21.750 / +60:51:41.420 by detecting a secondary X-ray source with coincident source detections in the \hst\  WFC3 H-band and Spitzer/IRAC images presented in \cite{Wang2024}. We observed the AGN in the field of the \xmm\ with the GNIRS instrument on the GEMINI-North observatory on February 21, 2024 using the long-slit mode (program ID: GN-2023B-FT-210). The slit has a length of $99^{\prime\prime}$. The observations were conducted with the AGN at the center of the slit using an ABBA dithering pattern. If the AGN is at the redshift of the HyLIRG, we expect to observe strong H$\beta$ emission. If the AGN is at a lower redshift than \xsc, we estimated that $\sim$ 5 hour integration time would be enough to recover signal from an AGN.

\par To reduce the GNIRS data, we used the GNIRS IRAF commands provided in the Gemini Virtual Machine (GemVM). We use the \texttt{nsprepare} command to align the spectra by correcting for any shifts in the slits during the observation and to apply a bad pixel mask. The regions outside the slit are cut using \texttt{nsreduce}. A flat field is generated using the \texttt{nsflat} command. An argon lamp with known emission line positions is used to measure any shifts in wavelengths during the observation. A telluric standard star, HIP 61371, is used to identify atmospheric absorption lines. Furthermore, we use the telluric star to determine the wavelength correction using \texttt{nsfitcoords} and the \texttt{nswavelength} command to apply the correction. A final combined spectrum is produced using \texttt{nscombine} and extracted using \texttt{nsextract}. The atmospheric absorption lines from telluric standard star are applied using \texttt{nstelluric}. The wavelengths in the final spectra that experienced high amounts of atmospheric contamination are removed from our analysis. 

We analyze the reduced spectrum by first fitting the continuum using the \texttt{fit\_generic\_continuum} command from \texttt{astropy} python package \texttt{specutils}. We identify emission lines above a S/N threshold of 3 using the \texttt{specutils} command \texttt{find\_lines\_threshold}. To measure a redshift, we fit the spectrum with a QSO template provided from \cite{Glikman2006}.

\section{\xmm\ Observations and Data Analysis}\label{s:xmm}

\subsection{\xmm\ observations}\label{ss:d-obs}
Most of our analysis focuses on the strong X-ray detection of \xsc. The reduction process of the \xmm\ data on \xsc\ is described in \cite{Wang2024}, but we provide a summary of the relevant details in this section. Briefly, data were collected from the three X-ray CCD cameras comprising the European Photon Imaging Camera (EPIC) \footnote{\url{https://www.cosmos.esa.int/web/xmm-newton/technical-details-epic}}. Two of the cameras are MOS (Metal Oxide Semiconductor) CCD arrays, while the other uses pn CCDs (referred to as pn data hereafter). The angular resolution of the EPIC pn is about 6.6$^{\prime\prime}$ (FWHM) or 15$^{\prime\prime}$ (80\% energy-encircled radius), while the MOS detectors have slightly better resolutions (e.g., 6$^{\prime\prime}$ FWHM).

\subsection{Spectral Data}\label{ss:d-spec}
In the case of high-$z$ X-ray observations, careful reduction of the data is necessary to produce usable spectra. We process the \ins\ data with the Science Analysis System (SAS) v21.0.0 software. First, we create MOS and pn event files using the \texttt{emproc} and \texttt{epproc} routines. In addition, we filter out time intervals with high background activity\footnote{\url{https://www.cosmos.esa.int/web/xmm-newton/sas-thread-epic-filterbackground}}. We detect discrete sources using the \texttt{edetect chain} script. We extract source and background spectra using \texttt{evselect} for the MOS1/2 and pn instruments. To generate response and effective area files, we use \texttt{rmfgen} and \texttt{arfgen}, respectively. Different observations of each instrument are combined using \texttt{epicspeccombine} with the observations of the two MOS instruments being combined because of their similar properties.

We determine the centroids of sources using the \texttt{edetect chain} output. This provides the extraction coordinates for \xsc\ at (RA / Dec. J2000) 10:53:22.08 / +60:51:43.2 and the location of any nearby contaminating sources that are removed. We use an on-source spectral extraction aperture radius of 20$^{\prime\prime}$ and a corresponding annulus background extraction region with an inner radius of 35$^{\prime\prime}$ and an outer radius of 175$^{\prime\prime}$. These aperture sizes ensure that we obtain an accurate background and source spectral representation.

\subsection{Spectral Modeling}\label{ss:d-spec-modeling}

Identifying possible sources of contamination is necessary to obtain accurate spectral results. In \cite{Wang2024}, we estimate the diffuse hot plasma contributions of the foreground lens and find them to be negligible (count rates $\lesssim 1\%$ in the 0.5-2~keV band and less in the broader bands).

\par The extracted spectra are grouped into seven counts per bin using the High Energy Astrophysics Science Archive Research Center (HEASARC) routine \texttt{ftgrouppha}. This grouping is recommended when using W-statistics (\texttt{wstat}), which is derived from C-statistics (\texttt{cstat}) and is suitable for data exhibiting Poisson statistics \citep{Kaastra2017}. We use the X-ray spectral fitting package \texttt{Xspec} \citep{Arnaud1996} to obtain an initial best fit of our spectra. We then input the data into the Bayesian X-ray Analysis (BXA) \citep{Buchner2014} software to perform our Monte Carlo Markov Chain (MCMC) analysis. To plot the results of the MCMC analysis we use the python package, chainconsumer \citep{Hinton2016}. Furthermore, the spectra plots are grouped to obtain an S / N $>$ 1 per bin to better visualize the data. 
\par To model our spectra in \texttt{Xspec}, we first account for Galactic absorption from the Milky Way with the \texttt{tbabs} component at a fixed column density of $9.6 \times 10^{19}$ cm$^{-2}$. The AGN is modeled using the \texttt{RefleX} AGN model \citep{Paltani2017}. The \texttt{RefleX} component \texttt{{RXTorus\_rprc}} models the reprocessed light by fluorescence and scattering. The AGN continuum is modeled using a cutoff power-law component, \texttt{clumin*({RXTorus\_cont}*zcutoffpl)}, with a cutoff turnover at restframe 200 keV. The convolution model, \texttt{clumin}, measures the luminosity in the restframe 0.5 - 8.0 keV band and normalizes each corresponding model component. Both components of the \texttt{RefleX} model measure distinct column densities. For the HMXB component, we use a simple absorbed power-law, \texttt{ztbabs*(const*clumin(zpowerlw))}. The HMXB luminosity is fixed to $6.69 \times 10^{43}$ ergs s$^{-1}$, which is the expected luminosity if \xp\ $= 1$ with the measured $\mu SFR$ of 16735 M$_{\odot}$ yr$^{-1}$ of \xsc\ \citep{Berman2022}. This allows the multiplicative constant component, \texttt{const}, to measure the \xp~(e.g. an \xp~of 2 corresponds to a measured luminosity twice that of the expected luminosity from the Mineo relation). The complete spectral model we use is described as follows.

\begin{equation}
    N_{H,Gal} (N_{H,HMXB}\times X_{HMXB}\times P_{HMXB} + R + C\times P_{cut})
\end{equation}
Here $N_{H,Gal}$ is the fixed Galactic absorption, $N_{H,HMXB}$ is the absorption from the HMXB population of \xsc, \xp\ is the parameter from Eq. \ref{e:xsfr}, $P_{HMXB}$ is the HMXB redshifted power-law, $R$ is the AGN reprocessed \texttt{RefleX} component, $C$ is the AGN continuum \texttt{RefleX} component, and $P_{cut}$ is the cutoff power-law. 

To limit the number of free parameters, we fix the AGN torus viewing angle to 90 degrees (edge on) and the inner-to-outer torus radius ratio (covering factor) of the torus to 0.5 (e.g. more ``donut"-shaped torus geometry) \citep{Ricci2021}. This produces a model with the following free parameters: non-AGN $N_H$ (cm$^{-2}$), non-AGN photon index, non-AGN luminosity \xp\ $(3.9 \times 10^{39} {\rm~erg~s^{-1}}/{\rm M_\odot~yr^{-1}})$, $\rm AGN_{\rm SW}$ torus column density $N_{H,t}$, $\rm AGN_{\rm SW}$ line-of-sight (LoS) $N_{H,LOS}$ (cm$^{-2}$), $\rm AGN_{\rm SW}$ photon index, and $\rm AGN_{\rm SW}$ luminosity (erg s$^{-1}$). We find that the \texttt{cstat}/d.o.f. ratio alone \footnote{unlike the $\chi^2$/d.o.f. ratio} cannot be used to judge if a fit is acceptable. To make such a judgment, we calculate the goodness of fit via a bootstrapping method, as described in the caption to Table \ref{t:spec} and in Appendix~\ref{a:sims}.

\subsection{Imaging Data}
The high number of counts observed in the X-ray images within a 2$^{\prime} \times$ 2$^{\prime}$ Field Of View (FOV) displayed elongated ellipsoidal emission with two peaks that have $>$ 30 counts located at the centroids of the southwest AGN and \xsc. This allowed us to perform a 2D spatial decomposition of the X-ray emission to independently model the amount of X-ray emission from $\rm AGN_{\rm SW}$ and \xsc. 
We use pipeline product (PPS) images in the 0.2-12.0 keV band. The centroids of the sources are identified using the \texttt{SAS} script \texttt{edetect\_chain}. We use the centroids of three bright sources, QSO SDSS J105315.15+605145.7, 4XMM J105415.9+60562, and 4XMM J105234.3+605654, to create a World Coordinate System (WCS) file with the \texttt{CIAO} script \texttt{wcs\_match} that matches the WCS of the observation with the highest exposure time (Obs. 0882720501). An astrometric correction is applied to the images of the remaining three observations [Obs. 0882720(201,701,901)] using the \texttt{CIAO} script \texttt{wcs\_update}. 

The corrected images of the three observations are reprojected to match the image coordinates of the 0882720501 exposure using \texttt{reproject\_image}. The resulting images are combined to produce one merged image. A Point-Spread Function (PSF) image is produced for each observation using the \texttt{SAS} script \texttt{psfgen}. This script accounts for various instrumental effects unique to each observation, such as roll and azimuthal angle. The PSF is energy-weighted as each PSF is composed of 81.7\% at 3.0 keV, 16.6\% at 6.0 keV, and 1.7\% at 9.0 keV energies. The weights are determined on the basis of the number of source counts in the 0.5-3.5, 3.5-7.0, and 7.0-10.0 keV bands. The produced PSFs of each observation are weighted based on the observation exposure time and combined to produce a merged PSF image. To confirm the accuracy of this merging process, we compared radial profiles for source QSO SDSS J105315.15 + 605145.7 (a bright quasar that is expected to be point-like) with the PSF using a single exposure image and the merged image (see Appendix~\ref{a:rad}). 

\begin{figure}
\centerline{
\includegraphics[width=1.0\linewidth,angle=0,scale=1.0]{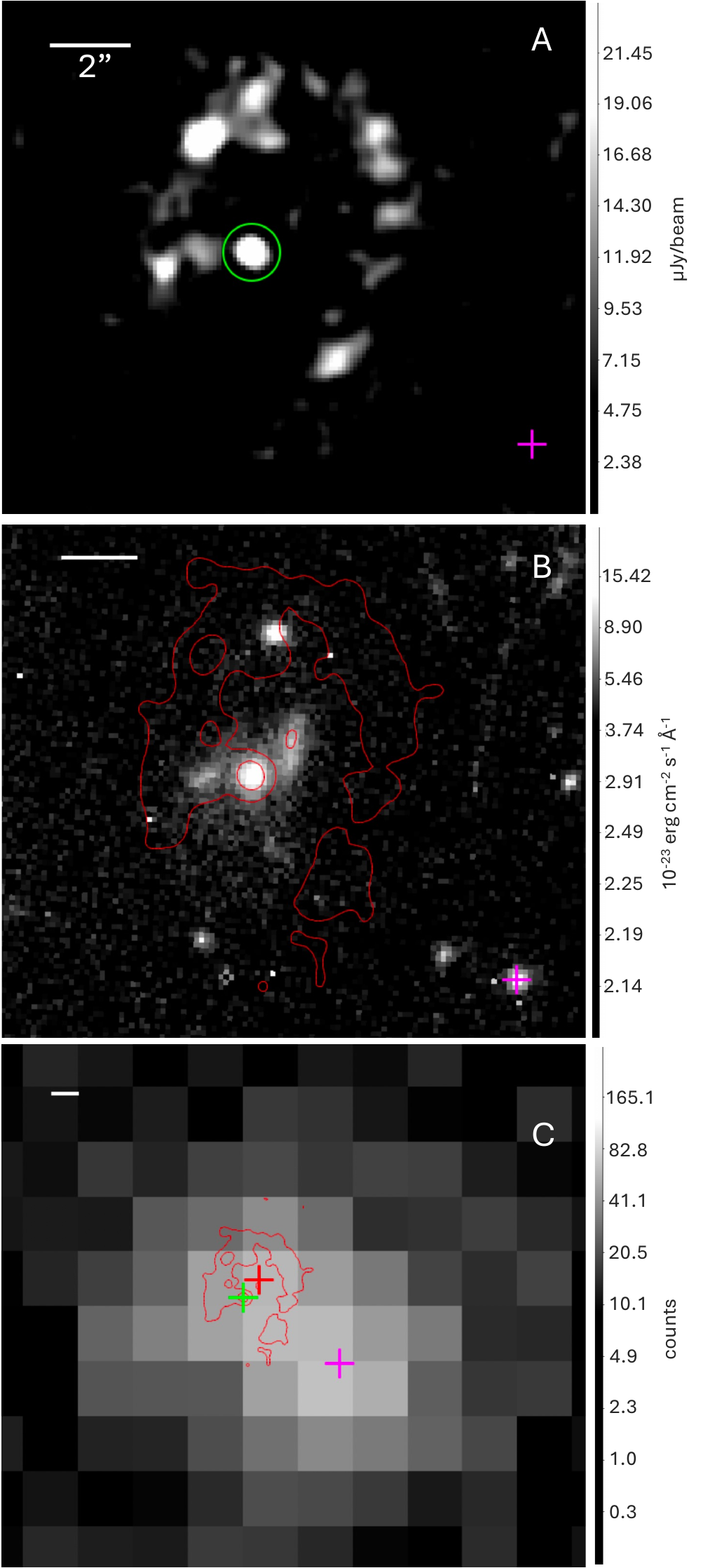}
}
\caption{Images comparing the different centroids used in our testing for a potential AGN in the foreground lens. The top left corner bars indicate $2^{\prime\prime}$ length. (A) VLA image showing \fg\ in the green circle and the magenta cross is the centroid of \sw. (B) \HST\ F160W image with VLA contour overlaid on the image. (C) X-ray image with different centroids and VLA contour overlaid. The green cross is the centroid of the \fg\, and the red cross is the centroid of the SMA image used in the 2D image modeling.}
\label{f:VLA}
\end{figure}

\begin{figure}
\centerline{
\includegraphics[width=1.0\linewidth,angle=0,scale=1.05]{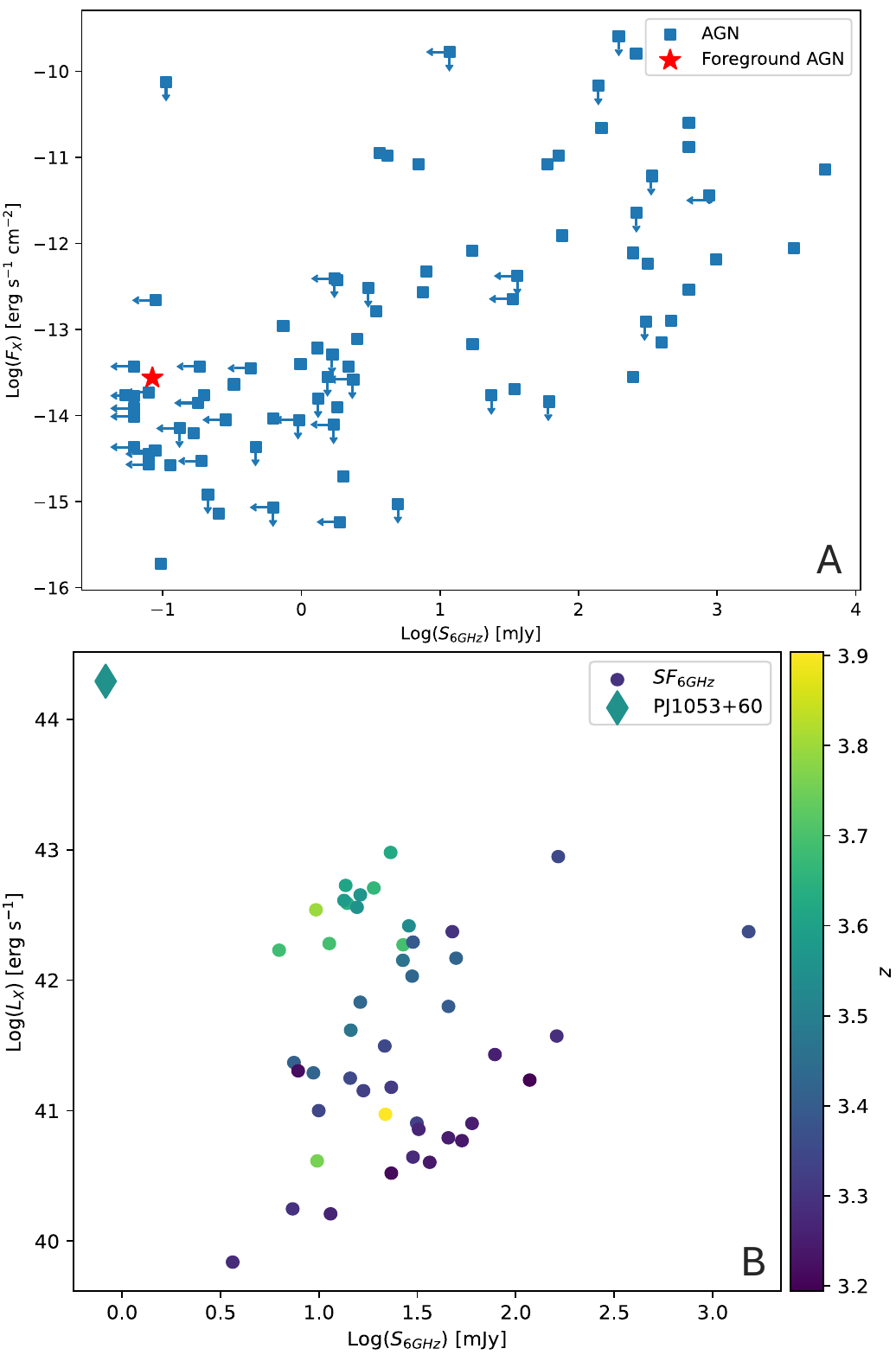}
}
\caption{Plots comparing X-ray flux and luminosities to 6 GHz flux (A) $F_X-S_{6 GHz}$ relation of the foreground AGN and the G{\"u}ltekin AGN sample. This shows that \fg\ exhibits similar $F_X$ and $S_{6 GHz}$ to other AGN. (B) Expected $L_X$ from $SFR$ and measured $S_{6 GHz}$ of \xsc\ compared to the expected $L_X$ from $SFR$ and measured $S_{6 GHz}$ from star-forming sources. Here, we assume \xp\ $\sim 3$ for any sources at $z\geq3$, including \xsc. The color of each marker is the redshift of each source to show that \xsc\ exhibits similar trends to other sources at high-z.}
\label{f:Fx}
\end{figure}

\begin{figure*}

\centerline{
\includegraphics[width=1.0\linewidth,angle=0,scale=1.0]{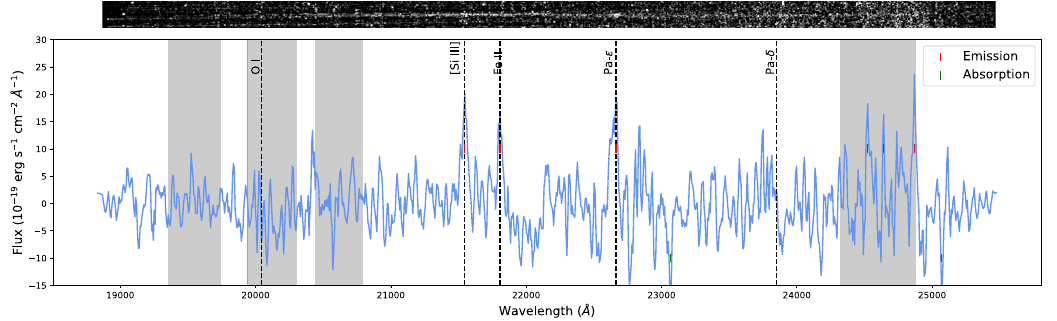}
}
\caption{Reduced continuum subtracted GNIRS spectra of the AGN with 2D spectra shown in the top panel. The gray regions indicate wavelengths contaminated with significant atmospheric absorption lines. The top panel shows the 2D image cutout of the spectra. Here three emission lines [Si III], Fe II, and Pa-$\epsilon$ have strong detections with S/N $>$ 3.}
\label{f:gnirs-spec}
\end{figure*}

\subsection{2D Spatial Modeling}\label{s:2dspat}

The main goal of modeling the spatial data of \xsc\ is to accurately represent the three main components of our target: the HMXB/$SFR$ X-ray emission, the southwest AGN counts, and the X-ray background. Given the physical separation of $\sim6.6^{\prime\prime}$ between the centroids of the southwest AGN and \xsc, good modeling is required to spatially decompose the two peaks of the ellipsoidal morphology of the X-ray emission. That in turn can lead to a measurement of \xp\ that is independent of redshift and can better handle degeneracies that arise from having to simultaneously model the southwest AGN and \xsc\ in the spectral analysis. Ultimately, the 2D image modeling enables us to better differentiate between AGN emission and HMXB emission.  

To model the HMXB component, we assume the distribution of dust traces the star formation in \xsc. We use the SMA dust continuum image from \cite{Wang2024}, ignoring noisy regions with a custom mask, and convert the SMA image into X-ray count rate images using the following relation.

\begin{equation}
    L_{x,i} = L_x \frac{I_i}{I}
\label{e:2d} 
\end{equation}

\par Here $L_{x,i}$ is the X-ray luminosity of each pixel, $L_{x}$ is the total expected X-ray luminosity of $6.69 \times 10^{43}$ erg s$^{-1}$, $I_i$ is the relative specific intensity of each pixel in the SMA image, and $I$ is the sum over all pixels. The luminosity value is converted to a count rate in cts s$^{-1}$ by entering $L_{x}$ into \texttt{XSPEC} using the HMXB simple power-law model (\texttt{tbabs * (ztbabs * clumin(zpowerlw))}) with an intrinsic column density of $10^{22}$ cm$^{-2}$ and a photon index of 2.0. The count rate of each pixel is converted using Eq.\ref{e:2d} to produce a count rate image that represents the expected X-ray spatial distribution of the lensed galaxy. Simply, we are assuming the HMXB X-ray luminosity follows the distribution of the lensed dust continuum seen in the SMA image. To produce an \xmm\ image model, we convolve the count rate image with the merged \xmm\ PSF. The convolved count rate image is reprojected using \texttt{reproject\_image} to match the pixel coordinates of the merged \xmm\ counts image (see Fig. ~\ref{f:2d-pro}), producing the final expected counts image model.

\par We use the final count rate image as the \xp\ component in the following Chandra Interactive Analysis of Observations (CIAO) \texttt{sherpa} model.
\begin{equation}
    M_{2D} = (X_{HMXB} + G_{AGN})E_{map} + C_{bkg}
\end{equation}
Here, $M_{2D}$ is the total 2D counts model. The $X_{HMXB}$ component represents the count rate image from \xsc, which is equivalent to the \xp\ component from \citet{Wang2024} and the \xp\ spectral model component described in \S~\ref{ss:d-spec-modeling}. The $G_{AGN}$ component, [\texttt{psf$\circledast$(Gauss2D.AGN)}], is a 2-D Gaussian model convolved with the \xmm\ PSF to represent the AGN emission. This Gaussian AGN model is the only component convolved with the PSF as the $X_{HMXB}$ component is previously convolved with the PSF in the process of producing the $X_{HMXB}$ expected count rate image. The AGN Gaussian is centered at a fixed position of RA 10:53:21.75 and Dec +60:51:41.4. 
The background counts, $C_{bkg}$, are modeled with a constant component, \texttt{Const2D.bkg}, because the $\sim$ 33 $\times$ 33 pixels region used to perform the 2-D fit is small. The $E_{map}$ is the exposure map image with each pixel value being the total exposure time of the combined observations for each pixel. This model contains 4 free parameters: $X_{HMXB}$ $(3.9 \times 10^{39} {\rm~erg~s^{-1}}/{\rm M_\odot~yr^{-1}})$, the Gaussian FWHM, $FWHM_{AGN}$ (pixel size of 4$^{\prime\prime}$), of the $\rm AGN_{\rm SW}$ component, the peak amplitude, $N_{AGN}$ (cts), of the $\rm AGN_{\rm SW}$ Gaussian component, and a constant background, $C_{bkg} (cts)$.

\begin{figure*}
\centerline{
\includegraphics[width=1.0\linewidth,angle=0,scale=1.0]{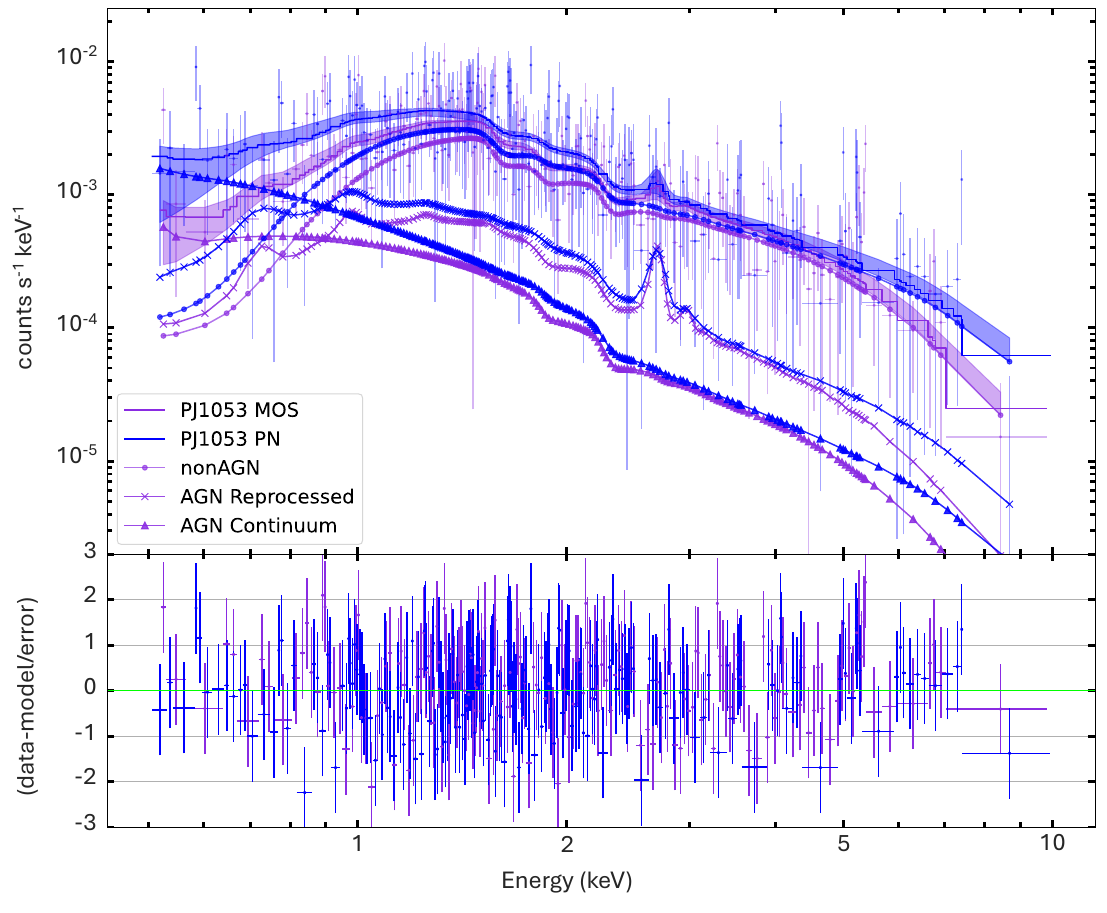}
}
\caption{\xmm\ spectral data with the best fit MCMC \texttt{RefleX} model (blue and purple shaded regions) at a 90\% confidence level. The data is split into the two \xmm\ instruments, MOS and pn. Here, the dot markers are the nonAGN or HMXB emission, the X markers are the reprocessed light of \sw\, and the triangle markers are the continuum like from \sw\. Although most data points are within 1 or 2 error levels, the model has a high goodness percentage of 86.7\% (see Table~\ref{t:spec}). This suggests that another AGN component from \fg\ is required to provide a better g.o.f.; however, due to the small spatial separation from \xsc\ and \fg\, we cannot disentangle the spectra into three components: \xsc, \sw, and \fg.
}
\label{f:x-spec}
\end{figure*}

\begin{figure*}
\centerline{
\includegraphics[width=1.0\linewidth,angle=0,scale=1.0]{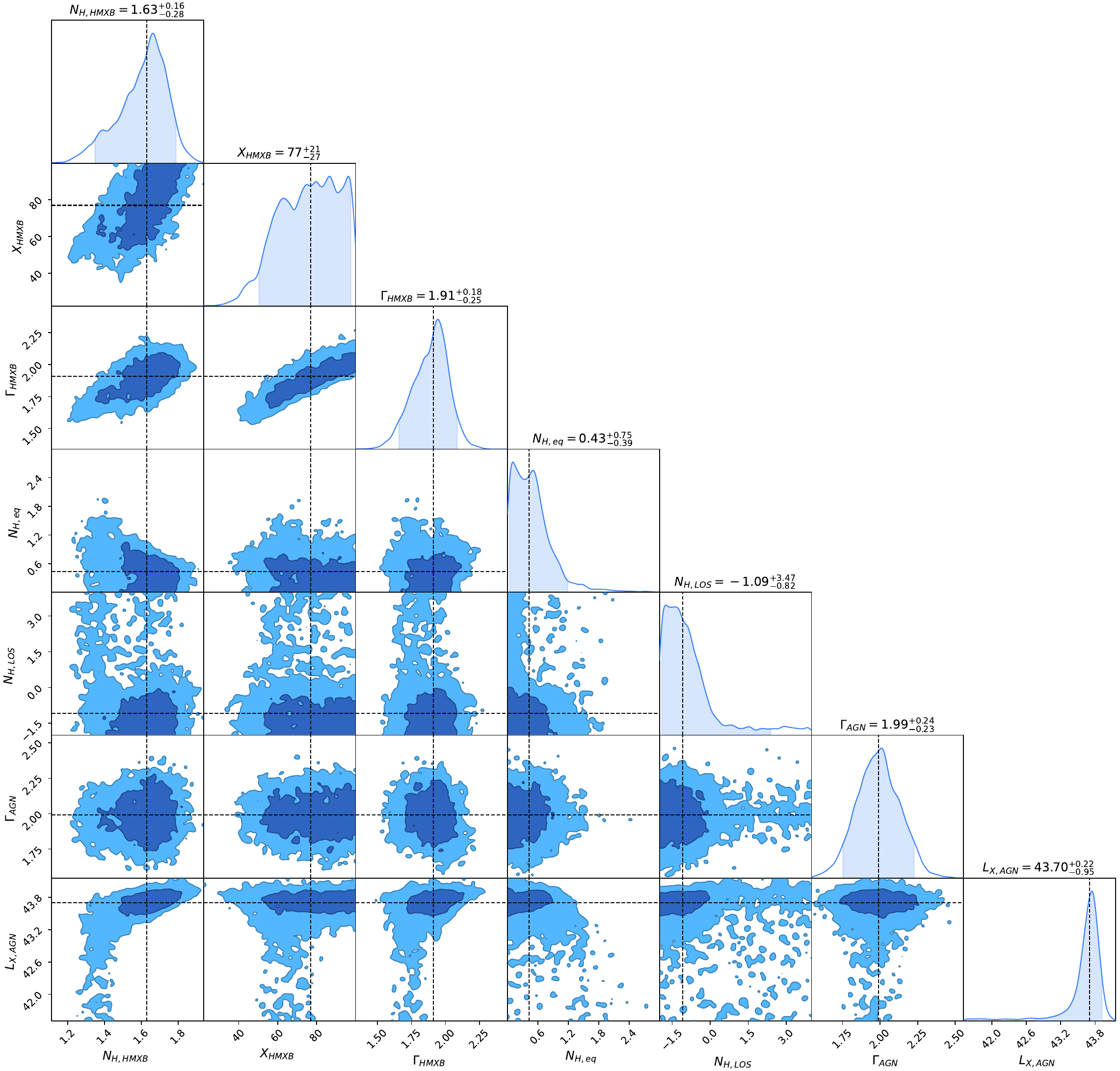}
}
\caption{Corner plot of the MCMC results using the \texttt{RefleX} model.
}
\label{f:corner}
\end{figure*}

\section{Results}\label{s:res}

\par Here, we present the results of our 2D spatial decomposition and the updated spectral modeling, as well as the redshift measurement of \sw. The main goal of this study is to obtain a reliable measurement of \xp\ by addressing any possible scenarios that may restrict our interpretation of the results. Throughout our analysis, the most significant limiting factor is the degeneracy between the HMXB counts and any contaminant AGN counts from $\rm AGN_{\rm SW}$ and $\rm AGN_{\rm FG}$. With the spectroscopic measurement of the southwest AGN redshift, we expect a better constraint on the \xp\ parameter using the \texttt{RefleX} spectral model compared to the value measured in \cite{Wang2024}. Furthermore, the 2D image modeling measures an \xp\ value independent of the redshift value and should spatially decompose the HMXB and $\rm AGN_{\rm SW}$ emission. However, potential degeneracies may still arise due to the number of counts in the spectra not being able to constrain a complex spectral model consisting of \fg, \sw, and \xp\ components, and the low spatial resolution X-ray image data used in our 2D image fit. In any case, we will discuss possible scenarios to assess their feasibility.  A summary of our results can be seen in Table \ref{t:spec}. We begin by addressing the possibility that there is an AGN embedded in the foreground lens. 

\begin{figure*}
\centerline{
\includegraphics[width=1.0\linewidth,angle=0]{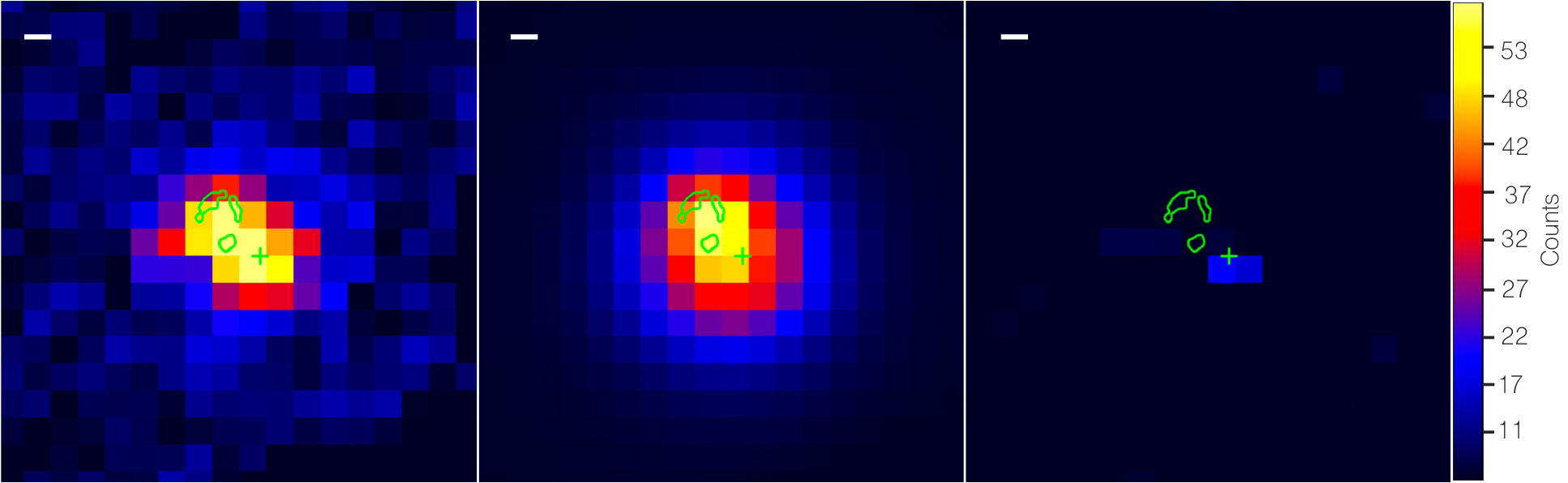}
}
\caption{The results of the 2D spatial fit to the \xmm\ image. The green contours show the masked SMA image referenced in Fig.~\ref{f:2d-pro}. The white bars in the top left corners represent $4^{\prime\prime}$. The green cross indicates the centroid of \sw. From left to right: the \xmm\ image, the best fit 2D model, and the residual. Here, the residual is the Image minus the Model. All images use the same color bar scaling and limits. The low counts seen in the residual suggest that our 2D model described in \S~\ref{s:2dspat} produces a satisfactory fit. }
\label{f:2D-fit}
\end{figure*}

\begin{table*}
\caption{Results of Spectral and Spatial analyses}\label{t:spec}
\vspace{-0.2cm}
{
\begin{tabular}{lcccccccccr}
\hline\hline
Fit & ReFlex & 2D Spatial \\
\hline
\texttt{wstat}/d.o.f & 672.70/695 & 2879.01/1020  \\
Goodness (\%) & 86.7 & 49.2  \\
$\Gamma_{HMXB}$ & $1.91 (1.66,2.09)$ &   \\
$X_{HMXB} (3.9 \times 10^{39} {\rm~erg~s^{-1}}/{\rm M_\odot~yr^{-1}})$ & $77 (50,98)$ & $28.57 (21.84,38.96)$\\
log[$N_{H,HMXB} (10^{22} {\rm~cm^{-2}}$)] & $1.63 (1.35,1.79)$ & \\
-$f_{X,HMXB} (10^{-15} {\rm~erg~s^{-1}~cm^{-2}})$ & $30.5 (22.4,39.9)$ & \\
$\Gamma_{AGN}$ & $1.99 (1.76,2.23)$ & \\
log[$N_{H,LoS} (10^{22} {\rm~cm^{-2}}$)] & $-1.09 (-1.98,2.38)$ & \\
log[$N_{H,t} (10^{22} {\rm~cm^{-2}}$)] & $0.43 (0.04,1.18)$ & \\
log[$L_{X,AGN} ({\rm~erg~s^{-1}})$] & $43.70 (42.75,43.92)$ &   \\
-$f_{X,AGN} (10^{-15} {\rm~erg~s^{-1}~cm^{-2}})$ & $2.2 (12.8,20.9)$ &   \\
$N_{AGN} (\rm cts)$ & & $51.65 (24.56,81.96)$\\
$FWHM_{AGN} (4^{\prime\prime})$ & & $4.68(3.67,6.88)$\\
$C_{bkg} (\rm cts)$ & & $6.15 (5.97,6.35)$\\
\hline
\end{tabular} 
}
\vskip -0.05cm

\noindent{The goodness value of a fit represents the percentage of 1000 data simulations that have a smaller \texttt{wstat}/d.o.f., while the 90\% confidence interval of each parameter is included in the parenthesis next to the best-fit value. The unit of \xp\ represents the $L_{X,HMXB}$ of the source. The \xp\ parameter is further described in  in Eq.~\ref{e:xsfr}. The AGN luminosity is calculated in the rest frame 0.5-8.0~keV band of each source, while the absorbed fluxes of the HMXB and AGN components ($f_{X,HMXB}$ and $f_{X,AGN}$) are inferred (hence the marked '-' ) from the best-fit model in the observed 0.5-8.0~keV band. 
}
\end{table*}


\subsection{AGN in the foreground lens}\label{s:dis}

A re-examination of the 6 GHz VLA image of \xsc\ presented in \cite{Wang2024} led us to speculate that the point-like source with a peak radio flux ($L_R$) of 0.055 mJy at the center of the foreground lensing group (see Fig.~\ref{f:VLA}) at $z=0.837$ might be an AGN. Additionally, the point-like source's location matches the BGG's position in the foreground galaxy group visible in the \hst\ image of the field. To test this hypothesis, we conduct a 2D model fit test on the X-ray image. The model consists of two 2D Gaussian components (representing two point-like sources) fixed at the coordinates of $\rm AGN_{\rm SW}$ and $\rm AGN_{\rm FG}$. This model gives a good fit with a goodness percentage of 47.8 \%, suggesting that the spatial distribution of the observed X-ray emission is consistent with an origin in the two AGNs. Similarly, we find that the observed \xmm\ spectrum can be fitted with a spectral model (see \S~\ref{ss:d-spec-modeling}) assuming that the emission is entirely due to the southwest $\rm AGN_{\rm SW}$ and the $\rm AGN_{\rm FG}$. From this spectral model (parameters referencing \fg\ are denoted with an $_{FG}$ subscript hereafter), we obtain a hydrogen column density of $N_{H,FG} = 2.75^{+1.82}_{-1.06} \times 10^{22}$ cm$^{-2}$, a photon index of $\Gamma_{FG}=1.87^{+0.22}_{-0.18}$, and a luminosity of $L_{X,FG}=1.00^{+0.19}_{-0.39} \times 10^{44}$ erg s$^{-1}$ or a flux at $z=0.837$ of $\sim2.91 \times 10^{-14}$ erg s$^{-1}$ cm$^{-2}$.  

With $L_{X,FG}$ we can estimate BH mass ($M_{BH}$) of $\rm AGN_{\rm FG}$ using the $L_{R}$ - $L_X$ - $M_{BH}$ fundamental plane relation from \cite{Gultekin2019}. We first convert the 6 GHz VLA peak flux to a 5 GHz flux with the relation: $S_{\nu} \propto \nu^{-\alpha}$ with $\alpha = 0.7$ \citep{calistro2017,Foord2024}. We then obtain an \fg\ mass of log($M_{BH,FG}/M_{\odot}) = 8.41_{-1.02}^{+1.06}$ (at a 1 $\sigma$ confidence level). This $M_{BH}$ value would be expected from an elliptical galaxy with a bulge mass of $\sim 10^{11} M_{\odot}$ \citep{Kormendy2013}. 
From the LTM lens model, we estimate this main lensing group has a total mass of $\sim7.7 \times 10^{12} M_{\odot}$ \citep{Wang2024}. 
Therefore, the potential \fg\ galaxy mass of $\sim 10^{11} M_{\odot}$ is a feasible estimate of the BGG mass of the main lensing group, suggesting that the X-ray emission in the direction of \xsc\ is consistent with arising entirely from \fg; however, based on the results seen in \cite{Wang2024}, we expect to still observe some X-ray emission from \xsc. 





The limited spatial resolution of the \xmm\ data and the presence of the two AGNs in close proximity to \xsc\ (see Fig.\ref{f:VLA}) make it difficult to quantify the non-AGN contribution from the HyLIRG. We therefore conduct a simple check to see how the $L_X$ from \fg\ of $L_{X,FG} = 1.00^{+0.19}_{-0.39} \times 10^{44}$ erg s$^{-1}$ and \fg\ BH mass estimate of log($M_{BH,FG}/M_{\odot}) = 8.41_{-1.02}^{+1.06}$ can change if the non-AGN or \xp\ contribution is included. 
Due to the lack of constraints on \xp, we assume an \xp\ value of $\sim 3$ -- the best high-$z$ estimate of \xp. This value is derived from the joint fit reported in \cite{Wang2024} of the other two PASSAGES \sous\ with \xmm\ observations, \xsa\ and \xsb. Notably, unlike \xsc, the VLA image of source \xsa\ shows no point-like features, suggesting that \xsa\ has no AGN contamination. Furthermore, the relatively low number of net X-ray counts of $\sim50$ are consistent with the $L_X$ from \xsa\ being entirely from HMXBs. Given the low net counts, a joint fit of \xsa\ and \xsb\ was performed where only the \xp\ parameters from each source are linked, producing our current best estimate of \xp\ $\sim 3$.

Using an \xp\ value of 3 and an apparent $\mu SFR \sim 16700$ ${\rm M_\odot~yr^{-1}}$ of \xsc, we can use Eq.~\ref{e:xsfr}~to find an expected apparent luminosity for \xsc\ of $\sim2.01 \times 10^{44}$ erg s$^{-1}$ or a flux at $z=3.549$ of $\sim1.67 \times 10^{-15}$ erg s$^{-1}$ cm$^{-2}$. Assuming that \xsc\ at an \xp\ $\sim 3$ contributes a value of $\sim1.67 \times 10^{-15}$ erg s$^{-1}$ cm$^{-2}$ to the total flux of \fg\ and \xsc, we measure a remaining flux of $\sim2.74 \times 10^{-14}$ erg s$^{-1}$ cm$^{-2}$ that we expect to be entirely from \fg. This result would mean that $94.3\%$ of the total flux of \fg\ and \xsc\ is comprised of flux from the foreground AGN, which is consistent with the expectation that \fg\ dominates the total flux. Subsequently, we find that including the HMXB contribution does not significantly change the fit (equally statistically satisfactory) with a BH mass of log($M_{BH,FG}/M_{\odot}) = 8.42_{-1.02}^{+1.06}$. To further assess the nature of the foreground AGN, we compare the X-ray flux of \fg, assuming \xp\ $\sim 3$, and $S_{6GHz}$ to the AGN sample in \cite{Gultekin2019} in Fig.~\ref{f:Fx}. We see that  \fg\ appears to be similar in X-ray and radio flux to the AGN sample from \cite{Gultekin2019}, despite the inclusion of \xp\ $\sim 3$. Furthermore, we assess the nature of \xsc\ by using the expected apparent X-ray luminosity from $\mu SFR$ of $\sim2.01 \times 10^{44}$ erg s$^{-1}$ and the measured $S_{6GHz} \sim 0.8$ mJy of \xsc\ and compare these values to the expected $L_X$ from $\mu SFR$ (using Eq.~\ref{e:xsfr}) and measured $S_{6GHz}$ from a sample of lensed star-forming sources in the Abell 2744 field. Here we only assume an \xp\ $\sim3$ for sources at $z\geq2$ given that this \xp\ value is only appropriate for high-z galaxies. Although \xsc\ exhibits the highest $\mu SFR$ of the sample with a value of $\sim16700$ ${\rm M_\odot~yr^{-1}}$, we find that \xsc\ seems to fall on the same trend of exhibiting a${\rm~log}(S_{6GHz}) \lesssim 1.5$ mJy and${\rm~log}(L_{X}) \gtrsim 42$ erg s$^{-1}$ as the other lensed sources at $z\sim3.5$. Ultimately, the plots seen in Fig.~\ref{f:Fx} display two main features: (a) that \fg\ exhibits consistent $L_{X,FG}$ and $S_{6GHz,FG}$ to other AGN even with exclusion of the HMXB contribution from \xsc\ with an \xp\ $\sim 3$ and (b) that the expected $L_X$ and measured $S_{6GHz}$ of \xsc\ with an \xp\ $\sim 3$ exhibits similar trends of expected $L_X$ from $\mu SFR$ and measured $S_{6GHz}$ of other lensed sources at $z\sim3.5$.
Therefore, we conclude that, although we are unable to fully quantify the HMXB contribution from \xsc\ due to the low X-ray spatial resolution, the scenario of \xp\ $\sim 3$ obtained in \cite{Wang2024} provides results that are consistent with other lensed star-forming sources like \xsc\ and other AGN like \fg.


\subsection{AGN Redshift}\label{s:gspec}
\par The reduced Gemini spectrum seen in Fig.~\ref{f:gnirs-spec}~shows three emission lines with S/N $>$ 3 which we identify as [Si III], Fe II, and Pa-$\epsilon$ at a best-fit redshift of $z_{AGN_{SW}}=1.373 \pm 0.006$ using the QSO template. Although strong sky absorption lines contaminate a few wavelength regions, our redshift measurement ignores any lines within these regions. One expected line, Pa-$\delta$, was affected by a cosmic ray, necessitating the removal of some pixels, which may have resulted in the lack of detection. However, the three lines appear to be strong and suggest that the redshift of the southwest AGN is not a member of the foreground galaxy group at $z=0.837$ or in our background target at $z=3.549$. Therefore, the southwest AGN is only a contaminant in our data, not an AGN within a few kpc of \xsc.



\subsection{Results of X-ray Spectral Modeling of HMXB and Southwest AGN Components}



With the new redshift of the southwest AGN, we examine the results of the X-ray spectral modeling of \xsc\ shown in Fig.~\ref {f:x-spec}~, Fig.~\ref {f:corner}~, and Table~\ref{t:spec}. The absorbing gas column density of the non-AGN or HMXB power law component is high with a value of $\sim 4.3 \times 10^{23}$ cm$^{-2}$. We approximate the dust extinction to be $\sim 194.57$ using the following relation from \cite{Gulver2009}: $N_H = 2.21 \times 10^{21} A_V$. This level of extinction would be very high as $A_V$ values for DSFGs at similar redshifts can typically range from 2 -- 5 \citep[e.g.,][]{Perez2023}, and is therefore consistent with the non-detection of \xsc\ in the existing \hst\ image. Another scenario is a potential AGN embedded in \xsc, which would likely have a dusty torus at the center of \xsc. Given the source's high intrinsic $SFR$ of $\sim 700$ ${\rm M_\odot~yr^{-1}}$, we expect dust to be well distributed throughout the galaxy. Additionally, the current SED reported in \cite{Berman2022} shows no clear evidence of hot dust dominating the mid-IR wavelengths that may suggest the presence of an AGN \citep{Kirkpatrick2012}.   
Further examining the other best fit parameters, the photon index, $1.91 (1.66,2.09)$, is flatter than what might be expected for emission originating from only HMXBs. This may suggest that the component may be significantly contaminated by $\rm AGN_{\rm FG}$ and may explain the apparent enhancement of \xp~$\sim$50 (see \S~\ref{s:dis}). However, due to the low spatial resolution of the X-ray data, we cannot conclusively rule out this possibility (see Fig.~\ref{f:x-spec}).

The lack of strong constraints on some parameters implies that the spectra do not have enough counts to assess the validity of a more complex model (see Fig.~\ref{f:corner}). To accurately incorporate another component like an additional AGN embedded in \xsc\ or an AGN part of the foreground lens, better quality spectra are needed. Although these results are suggestive, they are model-dependent in the spectral analysis. Independent confirmation and tests are therefore highly desirable. 

\subsection{2D Spatial Fitting Results}\label{s:2dres}

The 2D spatial modeling of the \xsc\ is thus a useful check (see Fig.~\ref{f:2D-fit}). Judging from the g.o.f. value  (49.2\%), the fit of the HMXB model appears to be satisfactory. 
With the constraint on its spatial extent FWHM= [4.68(3.67, 6.88) pixels], consistent with the PSF of the \xmm\ data, the southwest AGN component indeed appears to be point-like. However, with the limited spatial resolution of the data, the degeneracy between the southwest AGN and \xsc\ is still significant, leading to weak constraints on $N_{AGN}$ = [$51.65(24.56,81.96)$ counts] and \xp\ = [$28.57(21.84, 38.96)$ $(3.9 \times 10^{39} {\rm~erg~s^{-1}}/{\rm M_\odot~yr^{-1}})$]. 

Because the spectral modeling indicates a significant contribution from an additional AGN in the field of \xsc\, potentially in the foreground lens, we consider the possibility that this foreground AGN dominates the X-ray emission. This scenario is further explored in \S~\ref{s:dis}. However, we conclude that the \xmm\ data do not allow us to distinguish the morphology of the X-ray emission from the DSFG and that we cannot rule out the presence of another contaminating foreground AGN in the field.

\section{Summary and Conclusions}\label{s:con}

We have presented an improved analysis of the \xmm\ data on \xsc\ and a new redshift measurement based on a GNIRS spectrum confirming the presence of an AGN in the foreground lensing galaxy. Our main results are as follows.

\begin{itemize}
    \item Our spectroscopically measured redshift of the AGN is 1.373$\pm$0.006, suggesting that it is only a contaminant of \xsc, in between the lensing plane and the redshift of the DSFG. 
With the updated redshift of the AGN, we have improved the decomposition of the \xmm\ spectrum. Using a more sophisticated \texttt{RefleX} AGN model, we find the AGN has a luminosity of $\sim 5 \times 10^{43}$ erg s$^{-1}$. 
    \item Our spectral decomposition of the observed \xmm\ spectra into the southwest AGN and the \xsc\ contributions suggests that the latter would require an unusually high \xp\ value of $\sim$ 50 and a photon index too small to be consistent with what is expected from HMXBs (dominated by Ultraluminous X-ray sources) in an extreme star-forming galaxy at the rest frame energy above a few keV. This indicates that an additional AGN may contaminate the X-ray emission from this DSFG. Further evidence of this assessment is shown in our 2D spatial decomposition analysis of the \xmm\ data. 
    \item Our VLA data show the presence of a point-like radio continuum source that most likely represents an AGN in the central elliptical galaxy of the foreground lensing group. Using the $L_{R}$ - $L_X$ - $M_{BH}$ fundamental plane relation from \cite{Gultekin2019}, we measure a BH mass of the AGN of log($M_{BH,FG}/M_{\odot}) = 8.42_{-1.02}^{+1.06}$, which is consistent with what is expected from the mass of an elliptical galaxy in the foreground lens. 
    \item Although AGN contaminations have greatly complicated the measurement of $X_{HMXB}$, we show that its best fit value ($\sim3$), as obtained from the \xmm\ observations of the two other HyLIRGs \citep{Wang2024}, is consistent with the data on \xsc.
\end{itemize}


The above results, as well as those presented in \cite{Wang2024},  demonstrate the potential of X-ray observations in studying high energy phenomena of HyLIRGs and the limitations of presently available low resolution X-ray data.  In the \xsc\ case, high angular resolution X-ray observations are desirable to determine the population of HMXBs, which have important implications for understanding the $L_X - SFR $ relation and its relation to the extreme star formation in \sous.



\section*{Acknowledgment}
This work is supported by NASA under grant 80NSSC22K1923, based on data obtained from \xmm.
This paper also makes use of the following ALMA data: ADS/JAO.ALMA \# 2017.1.01214.S, as well as VLA (project ID: 18A-399) and SMA (observation ID: 7200) data. 
 The National Radio Astronomy Observatory is a facility of the National Science Foundation operated under cooperative agreement by Associated Universities, Inc.
The Submillimeter Array is a joint project between the Smithsonian Astrophysical Observatory and the Academia Sinica Institute of Astronomy and Astrophysics and is funded by the Smithsonian Institution and the Academia Sinica. E.F.-J.A. acknowledges support from UNAM-PAPIIT project IA102023, and from CONAHCyT Ciencia de Frontera (project ID: CF-2023-I-506). E.F.-J.A. acknowledge support from UNAM-PAPIIT projects IA102023 and IA104725, and from CONAHCyT Ciencia de Frontera project ID: CF-2023-I-506. 

\section*{DATA AVAILABILITY}

The X-ray data presented here are available in the \xmm\ data archive (https://www.cosmos.esa.int/web/xmm-newton/xsa). Other processed data products underlying this article will be shared on reasonable request to the authors.
\bibliographystyle{mnras}
\bibliography{export-bibtex} 

\appendix 
\section{Radial Profile}\label{a:rad}

\begin{figure}
\centerline{
\includegraphics[width=1.0\linewidth,angle=0,scale=1.15]{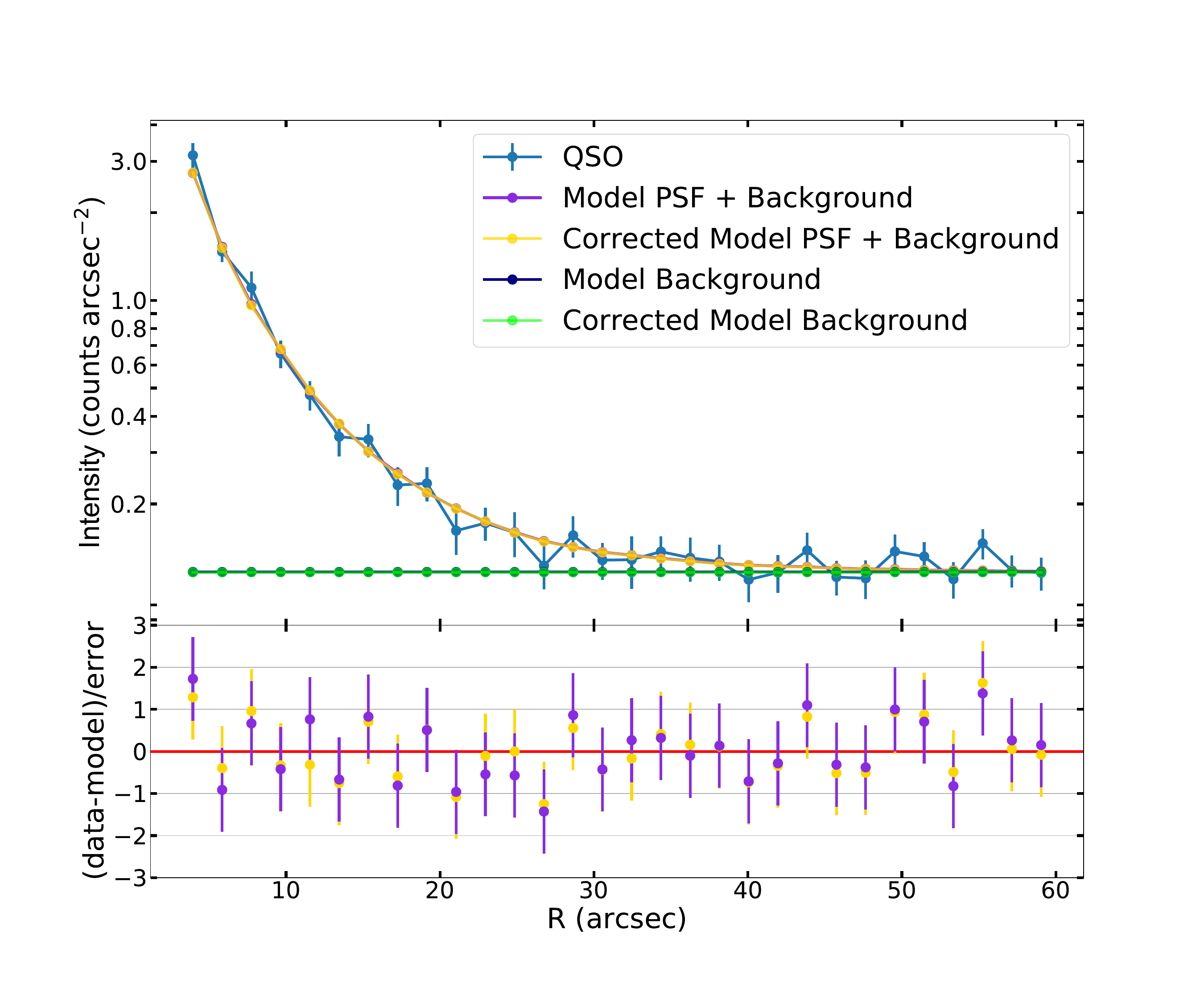}
}
\caption{Comparison of the \xmm\ PSF with the radial intensity profiles of the pn 0.5-4.5~keV emission for QSO SDSS J105315.15+605145.7. The plot markers labeled as "corrected" refer to the best fit of the PSF model after the astrometric correction. Although the corrected and uncorrected models are similar, the astrometric correction improves PSF model fit of the point-like QSO with a $\chi^2$/d.o.f. of 14.51/29.
}
\label{f:psf}
\end{figure}
\par To test the image merging procedure used in the 2D fit, we perform a radial profile PSF fit of QSO SDSS J105315.15+605145.7 using various X-ray datasets: a single observation image, a merged image, and the final merged image with astrometric correction. We remove any sources that may contaminate the radial profile. We create 30 annuli regions with an inner-most radius of 3 arcseconds and an outer-most radius of 60 arcseconds. The centroid used was R.A. 10:53:22.08 and Dec +60:51:43.2 from the \texttt{edetect chain} results. To estimate the background count rate we use an annulus region with radii of 60 and 75 arcseconds. The \texttt{CIAO} script \texttt{dmextract} is used to extract counts at each annulus producing a radial profile from background subtracted counts. The radial profile was fit using a simple PSF model described as follows.

\begin{equation}
    M_{PSF} = A \times R_{PSF} + B
\end{equation}
Here $A$ is the normalization, $B$ is the model background, and $R_{PSF}$ is the PSF. This model was fit to the radial profiles using $\chi^2$ statistics: 
\begin{equation}
    \chi^2 = \sum \frac{(O - M)^2}{\sigma^2}
\end{equation}
Here $O$ is the observed data, $M$ is the PSF model, and $\sigma$ are the uncertainties of each of the observed data points. In the case that the radial profile exhibits point-like behavior, it is expected to show a low $\chi^2$ value using this PSF model.

\par The initial fit using only the 0882720501 exposure resulted in a $\chi^2$/d.o.f. of 16.24/29, suggesting the QSO is point-like (see Fig.~\ref{f:psf}). The merged images radial profiles are conducted without the 201 exposure, as a chip gap covers $\sim$ 1/2 of the QSO in that exposure. The merged dataset results in an increase in the $\chi^2$/d.o.f. compared to the single exposure radial profile with a value of 17.97/29, suggesting that the QSO is point-like, but inaccurate merging of observations may result in a decrease of sharpness in the merged image. The final merged dataset with the astrometric correction produces a $\chi^2$/d.o.f. of 14.51/29, which is the best g.o.f. Thus, the astrometric corrected data produce a sharper image with better fitting statistics compared to both the single observation and the uncorrected merged image radial profiles. Subsequently, applying the astrometric correction results in a more accurate representation of sources compared to the uncorrected dataset and a single observation

\section{Goodness of Fit}\label{a:sims}

\begin{figure}
\centerline{
\includegraphics[width=1.0\linewidth,angle=0, scale=1.05]{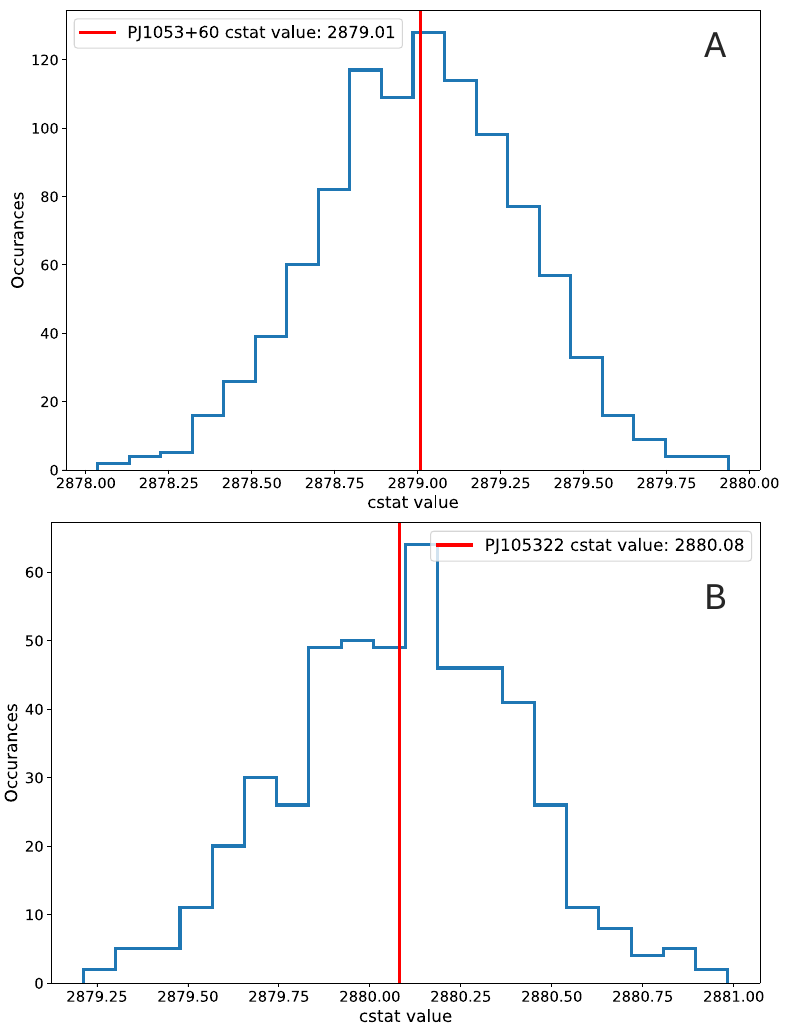}
}
\caption{Distribution of \texttt{cstat} values produced by 1000 simulations of the best fit model. (A) Histogram of \texttt{cstat} values for simulations of a simple 2D Gaussian model fit for QSO SDSS J105315.15+605145.7. (B) Plot of the simulations for \xsc. Because the simulations are randomly generated from the best fit model, a good fit would depict a normal distribution centered around the original best fit \texttt{cstat} value. }
\label{f:2D-sim}
\end{figure}

\par The \texttt{cstat} statistic is a likelihood fitting metric used to find the best fit of data exhibiting Poisson counting statistics. The general equation used in the \texttt{Sherpa} software is described as follows:
\begin{equation}
    C = 2 \sum_{i}[M_i - D_i ({\rm log}D_i - {\rm log}M_i)]
\end{equation}
where $M_i$ is the sum of the source and background model counts ($M_i = S_i + B_i$) and $D_i$ is the observed counts in each spectral bin $i$. This fitting method depends critically on separately characterizing both the source counts and background counts. Therefore, it is often recommended to slightly bin the data to have $\gtrsim 5$ counts per bin \citep{Kaastra2017}. This is typically easy to execute in spectral data, however, in our case, we do not perform any binning of the images to maintain the spatial information of this system. Consequently, this can highly skew the \texttt{cstat} value in the 2D image fit, resulting in the \texttt{cstat}/d.o.f. value being unreliable. The \texttt{cstat} value is therefore best used only as a maximum likelihood statistic to measure parameter uncertainties and compare models.

In order to accurately assess the g.o.f. of the 2-D fit results we simulate datasets as a method to gauge the validity of the \texttt{cstat} values. When we introduce noise to the best-fit image model, the resulting image should still accurately represent the model but also include simulated randomness. Using the \texttt{python} script \texttt{skimage}, Poisson noise was added to the model image and refit using the same model and fit procedure. The best fit \texttt{cstat} value is then recorded with this process repeated 1000 times to obtain a distribution of 1000 \texttt{cstat} values. If the initial \texttt{cstat} value is a good fit, the resulting distribution of the \texttt{cstat} values should be randomly distributed and follow a normal distribution centered at the initial \texttt{cstat} value. This ultimately produces a goodness test percentage based on the number of simulated datasets that presented a lower \texttt{cstat} value that indicate a better fit. A goodness percentage of $\sim$ 50\% indicates randomly distributed \texttt{cstat} values or a satisfactory fit.

\par Using QSO SDSS J105315.15+605145.7 as a test, we perform the 2D goodness test using a simple 2D Gaussian model convolved with the \xmm\  PSF and a constant background component: \texttt{psf$\circledast$(Gauss2D.AGN))*emap + Const2D.bkg}. This model should be an accurate representation of a point-like source. The distribution of the 1000 simulated datasets showed a clear normal distribution (see Fig.~\ref{f:2D-sim}) with a goodness percentage of 50.8\%. Therefore, this test suggests that the current goodness test can produce the expected \texttt{cstat} value distribution from random noise. Conducting this goodness test on the fixed AGN centroid fit of \xsc\ produced a satisfactory percentage of 49.2\% with a consistent normal \texttt{cstat} distribution (see Fig.~\ref{f:2D-sim}). Ultimately, the 1000 \texttt{cstat} values appear consistent with randomly distributed \texttt{cstat} values with a goodness percentage of 49.2\% suggesting a satisfactory fit.

\end{document}